\documentstyle[epsf,prl,aps,twocolumn]{revtex}
\begin{document}
\draft
\title{Nonequilibrium transport for crossed Luttinger liquids} 
\author{Andrei Komnik and Reinhold Egger}
\address{Fakult\"at f\"ur Physik, Albert-Ludwigs-Universit\"at,
Hermann-Herder-Stra{\ss}e 3, D-79104 Freiburg, Germany }
\date{February 11, 1998} 
\maketitle
\begin{abstract}
Transport through two one-dimensional interacting metals 
(Luttinger liquids) coupled together at a single point is analyzed. 
The dominant coupling mechanism is shown to be of 
electrostatic nature. Describing the voltage
sources by boundary conditions 
then allows for the full solution of the transport
problem.  For weak Coulomb interactions, transport is
unperturbed by the coupling. In contrast, for
strong interactions, unusual nonlinear  
conductance laws characteristic for the correlated system
can be observed.
\end{abstract}

\narrowtext

The physics of one-dimensional (1D) conductors has received much
attention lately, chiefly due to fabrication 
advances and the discovery of novel 1D materials
such as carbon nanotubes \cite{tube}.  From the theoretical
point of view, these systems are of interest since
Coulomb interactions invalidate the ubiquitous 
Fermi liquid description.
The resulting state is often of Luttinger liquid (LL) \cite{voit,gogolin}
type characterized by, e.g.,
spin-charge separation, suppression of the tunneling density
of states, and interaction-dependent power laws in the
transport behavior. However, so far the unambiguous
experimental observation of  LL behavior
has been difficult to achieve.

In this paper, we study two correlated 1D metals coupled in 
a point-like manner (``crossed Luttinger liquids''). For
the standard two-chain problem, 
where two Luttinger liquids are connected all along the conductors,
the coupling normally destroys the LL phase
\cite{fabrizio}. In the case of a point-like coupling, however,
the LL characteristics can survive and lead to the
unusual transport features reported below. The most promising
candidates for their experimental observation 
are carbon nanotubes.  At not exceedingly low temperatures,
metallic single-wall nanotubes exhibit LL behavior
(with an  additional flavor index) \cite{egger97}.
In a remarkable recent experiment,  Tans {\em et al.} \cite{tans}
were able to attach leads to a single nanotube. So far, 
transport measurements have been
 dominated by Coulomb charging effects due
to rather large contact resistances between the leads and the
nanotube, thereby masking any possible deviation from Fermi liquid
theory.  In the near future this problem might be overcome, and 
non-Fermi liquid laws should indeed emerge.
Other realizations of crossed Luttinger
 liquids could be based on, e.g.,
1D quantum wires in semiconductor heterostructures \cite{wire},
or edge states in a fractional quantum Hall bar \cite{fqh}.

\begin{figure}
\hfil
\epsfysize=6cm
\epsffile{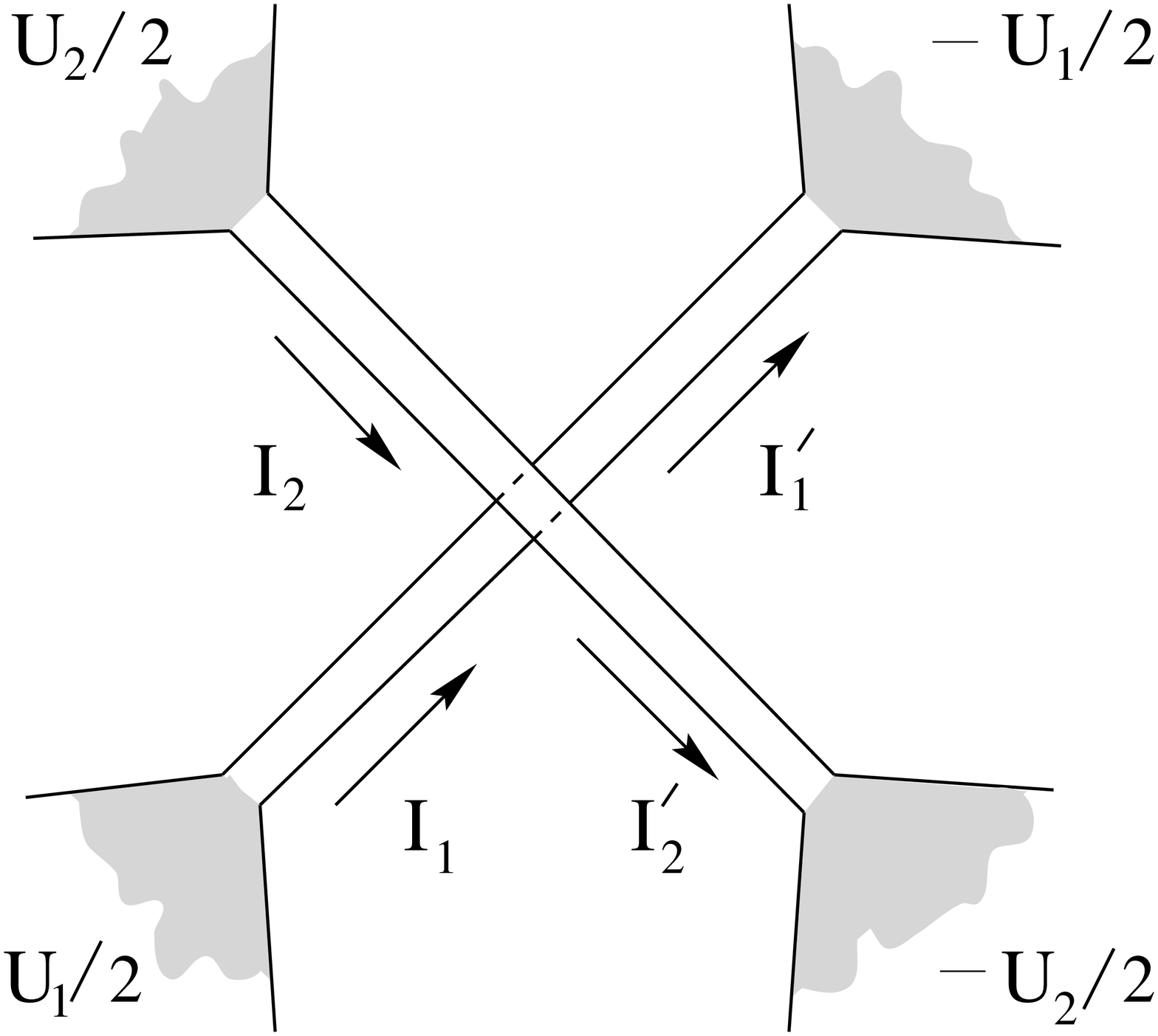}
\hfil
\caption[]{\label{fig1}
Two Luttinger liquids coupled together at one point ($x=0$)
and connected to external reservoirs
held at constant voltages $\pm U_1/2$ and $\pm U_2/2$.
In the absence of single-particle tunneling,
the currents obey $I^{}_i=I_i^\prime$.}
\end{figure}

The geometry of our system is shown in Fig.~\ref{fig1}.
We shall consider two spinless Luttinger liquids characterized
by the same interaction constant $g$ \cite{unp}.
 Here the noninteracting
value is $g=1$, and externally screened Coulomb interactions imply 
$0<g<1$ \cite{voit,gogolin}.
For the nanotube experiment of Ref.\cite{tans}, one has an 
externally unscreened $e^2/|x-x'|$ interaction potential and
therefore very strong correlations.
Strictly speaking, this interaction leads to $g\to 0$
in an infinite system, but the finite length of the nanotube in 
Ref.\cite{tans} implies $g\approx 0.2$.  
A natural and quite simple
description of Luttinger liquids is offered by the standard bosonization
method \cite{gogolin}.  External reservoirs (voltage sources)
can be incorporated by imposing boundary 
conditions \cite{egger96} for the phase fields 
employed in the bosonization scheme.
This approach offers a general and 
powerful route to studying multi-terminal Landauer-B\"uttiker
geometries \cite{datta} for strongly correlated electrons.
The crossed Luttinger liquids depicted in Fig.~\ref{fig1}
 may be the simplest example for such a problem.

We start by expressing the right- and left-moving ($p=\pm$)
component of the electron operator $\psi_{pi}(x)$
in conductor $i=1$ or 2  in terms of the dual
bosonic phase fields $\theta_i(x)$ and $\phi_i(x)$
obeying the algebra
\begin{equation}\label{alg1}
[ \phi_i(x), \theta_j(y)]_-= -(i/2) \delta_{ij} \,{\rm sgn}(x-y) \;.
\end{equation}
The bosonization formula then reads \cite{gogolin}
\begin{equation}\label{bos}
\psi_{pi} (x)= \frac{\eta_{pi}}{\sqrt{2\pi a}} \exp[-ipk_F x -
ip\sqrt{\pi g} \,\theta_i(x) - i\sqrt{\pi/g} \,\phi_i(x)] \;,
\end{equation}
where the same average density
$k_F/\pi$ is assumed for  both conductors. The short-distance
cutoff (lattice spacing) in Eq.~(\ref{bos}) is taken as $a=1/k_F$. 
To ensure anticommutation relations among different
branches $(p,i)$, we use (real) Majorana fermions $\eta_{pi}$ fulfilling
$[ \eta_{pi},\eta_{p'i'}]_+ = 2 \delta_{pp'} \delta_{ii'}$.
In the following, only products of Majorana fermions will
appear.  A valid choice for these products employs
 the standard Pauli matrices  \cite{egger97},
\begin{eqnarray} \label{majorana}
&& \eta_{p1}\eta_{p2} =i\sigma_x
\;, \quad \eta_{p1}\eta_{-p,2} = -ip\sigma_y\;, \\
\nonumber
&&\eta_{p1}\eta_{-p1} = ip\sigma_z \;, \quad \eta_{p2}\eta_{-p2}
= -ip\sigma_z\;. 
\end{eqnarray}
Assuming that the conductors
do not contain impurities, the
Hamiltonian of the uncoupled system is
\begin{equation}\label{h0}
H_0=\frac{1}{2} \int dx \sum_{i=1,2} \left\{ (\partial_x\phi_i)^2+
(\partial_x \theta_i)^2 \right\} \;,
\end{equation}
where we have put $\hbar=1$ and the sound velocity $v=v_F/g=1$.
Adiabatically connected voltage sources can then be taken into account 
by  Sommerfeld-like boundary conditions.
Applying the voltage $U_1$ along
conductor 1, and $U_2$ along conductor 2, 
see Fig.~\ref{fig1}, they read \cite{egger96}
\begin{equation}\label{bc}
\langle\rho_{p=\pm,i} (x\to \mp \infty)\rangle
=\pm \frac{eU_i}{4\pi g} \;,
\end{equation}
where $\rho_{\pm,i}$ is the density of right/left-moving particles
injected into conductor $i$. Outgoing particles are assumed to enter
the reservoirs without reflection.

Let us now consider a point-like coupling of both 1D conductors at,
say, $x=0$. For example, in a nanotube setup, 
two nanotubes could be stacked on top of each other. 
Such a contact causes (at least) two different coupling
mechanisms \cite{fo1}.

First, there arises a (density-density) 
{\em electrostatic interaction} of the form
$V_1= \lambda_1 \rho_1(0) \rho_2(0)$.  Using Eqs.~(\ref{bos})
and (\ref{majorana}),
and omitting the mean density $k_F/\pi$ which is supposedly
neutralized by positive background charges, the bosonized
representation of the density operator is
\begin{equation}\label{dens}
\rho_i(x) = \sqrt{\frac{g}{\pi}}\, \partial_x \theta_i (x) 
\mp \frac{\sigma_z}{\pi a} \sin[2k_F x+\sqrt{4\pi g}\, \theta_i(x)]\;,
\end{equation}
where the first (``slow'') term is due to the 
sum of right- and left-moving
densities $\rho_{Ri}+\rho_{Li}$, and the second (``fast'') term
arises from mixing right- and left-movers. 
The $\mp$ signs correspond to $i=1,2$, respectively.
One checks easily that most contributions to $V_1$
are irrelevant for $g\leq 1$, i.e., they have
scaling dimension $\eta>1$. Keeping  the
fast component in Eq.~(\ref{dens})  yields the only important term,
\begin{equation}\label{v1}
V_1 = -\frac{\lambda_1}{(\pi a)^2} \sin[\sqrt{4\pi g} \,\theta_1(0)]
\sin[\sqrt{4\pi g} \,\theta_2(0)] \;,
\end{equation}
with scaling dimension $\eta_1=2g$. Clearly, this coupling
becomes relevant for sufficiently strong interactions, $g<1/2$.
In contrast, a static potential scatterer in  
one of the conductors would be relevant already for 
$g<1$ \cite{kane}. In our case, electrons in conductor 1 experience
the {\em fluctuating} potential scattering $V_1$ due to electrons in
conductor 2, implying a doubled scaling dimension.

The second potentially important process is  
{\em single-particle hopping} from one 
conductor into the other. It is helpful
to distinguish processes that do (do not)
preserve the $p=R,L=\pm$ index, yielding the two perturbations
$V_2 =\lambda_2 \sum_{p} \psi^\dagger_{p1}(0) \psi^{}_{p2}(0)
+$~H.c.~(preserving the $p$ index) and
$V_3 =\lambda_3 \sum_{p} \psi^\dagger_{p1}(0) \psi^{}_{-p2}(0)
+$~H.c.~(not preserving the $p$ index).
They have the  bosonized form 
\begin{eqnarray} \label{v2}
V_2&=& - \frac{2\lambda_2}{\pi a} \sigma_x
\cos\{\sqrt{\pi g}\, [ \theta_1(0)-\theta_2(0) ] \} \\ 
\nonumber && \quad \times
\sin\{\sqrt{\pi/g}\, [ \phi_1(0)-\phi_2(0) ] \} \\
\label{v3}
V_3 &=& \frac{2\lambda_3}{\pi a} \sigma_y
\sin \{ \sqrt{\pi g}\, [\theta_1(0)+\theta_2(0)] \}
\\ \nonumber && \quad \times
\cos\{ \sqrt{\pi/g}\, [ \phi_1(0)-\phi_2(0) ] \} \;.
\end{eqnarray}
For the standard two-chain problem, a (bulk) coupling term 
formally identical to Eq.~(\ref{v2}) has been
discussed in Ref.\cite{kusmartsev}.
Both $V_2$ and $V_3$ have scaling dimension 
$\eta_2=\eta_3=(g+1/g)/2>1$,
from which one might naively conclude that they are  irrelevant
 \cite{foot}.
However, this conclusion is premature because  $V_2$ and $V_3$
have conformal spin $S=1$ \cite{fo2}.
For an operator with non-zero conformal spin, the
standard criterion for relevance $\eta<1$ does not apply, since
 relevant perturbations may be generated in higher orders of the
renormalization group (RG).
This phenomenon indeed occurs in the standard two-chain problem 
\cite{relevant},  where the (bulk) coupling term 
corresponding to Eq.~(\ref{v2}) generates
relevant particle-hole and/or particle-particle excitation
 operators. 
Similarly, we find that $V_2$ and $V_3$  together 
generate the electrostatic coupling $V_1$ given in Eq.~(\ref{v1}),
but no other relevant terms.  Omitting irrelevant operators,
the resulting RG equations take the closed form
\begin{eqnarray}\label{vflow}
\frac{d\lambda}{d\ell}  &=& [1-2g]\,\lambda + [g-1/g]\, t^2 \;, \\
\nonumber
\frac{dt}{d\ell} &=& [1-(g+1/g)/2]\, t \;,
\end{eqnarray}
where $\lambda_2=\lambda_3\equiv t$ is the
hopping amplitude  and $\lambda_1\equiv\lambda$
the electrostatic coupling.
The standard flow parameter is defined
by $d\ell=-d\ln \omega_c$, where $\omega_c$
is a high-frequency cutoff that is
reduced under the RG transformation.

Let us first discuss the case $g=1$.
The electrostatic coupling is irrelevant, i.e.,
 we may effectively put $\lambda=0$, but the
hopping term stays marginal. By refermionizing the 
Hamiltonian $H_0+V_2+V_3$, and employing the
 boundary conditions (\ref{bc}),
one arrives at the familiar results for  uncorrelated
electrons in the geometry of Fig.~\ref{fig1}, see Ref.~\cite{datta}.  
We therefore recover the usual Landauer-B\"uttiker formalism. 
Second, for $g<1$,  the hopping
amplitude $t(\ell)$ always scales to zero as $\ell\to \infty$, 
and the effects of single-particle tunneling can be 
captured by a renormalization of the bare electrostatic
coupling $\lambda$, see Eq.~(\ref{vflow}).
  We shall assume henceforth
that this renormalization has  been carried out,
and only the electrostatic interaction $V_1$ will be kept.
In that case, the currents flowing through conductor
$i=1$ or 2 satisfy
$I_i^{}=I_i^\prime$, see Fig.~\ref{fig1}, and can be
computed from the bosonized current operator  \cite{gogolin}
\begin{equation}\label{boscurr}
I_i= e\sqrt{g/\pi}\, \partial_t \theta_i(x=0,t)\;.
\end{equation}

For {\em weak interactions}, $1/2<g<1$, the
electrostatic coupling $\lambda(\ell)$ also flows to
zero as $\ell\to \infty$.  In that case, at low energy scales,
crossed Luttinger liquids are basically {\em insensitive} 
to the coupling considered here.
At asymptotically low energy scales,
the currents are then  $I_i=(e^2/h) \, U_i$.  
The finite-temperature or low-voltage
corrections due to the irrelevant operators $V_i$
 can be computed by  perturbation theory in the respective
coupling strengths $\lambda_i$.  Since the fluctuating potential scattering
is irrelevant for $1/2<g<1$, the corrections due to $V_{2,3}$  
are governed by the standard exponent 
$\alpha=(g+g^{-1}-2)/4$ for tunneling into a bulk
LL \cite{voit,gogolin}. This is in contrast
to a static potential scatterer, where tunneling into
the end of a LL matters at low energy scales \cite{kane}.

Directly at $g=1/2$, the operator $V_1$ is marginal,
and straightforward refermionization yields 
\begin{equation}\label{i2}
I_i = \frac{e^2}{h} \, \frac{1}{1+(\lambda/2\pi a)^2}\, U_i\;.
\end{equation}
Each conductor exhibits a response only
to the voltage applied to itself, with the conductance now explicitly
depending on the electrostatic coupling strength $\lambda$.
 
For sufficiently {\em strong interactions},
$g < 1/2$, the electrostatic coupling $\lambda(\ell)$ flows
 to strong coupling. To proceed,
we switch to the linear combinations
\begin{eqnarray}\label{newbas}
\vartheta_\pm (x) & =& \{ \theta_1(x) \pm \theta_2(x) \}
/\sqrt{2} \;,\\ \nonumber
\varphi_\pm (x) &=& \{ \phi_1(x) \pm \phi_2(x) \}
/\sqrt{2} \;,
\end{eqnarray}
which again obey the algebra (\ref{alg1}).
Remarkably, the Hamiltonian $H_0+V_1$ decouples into
the sum $H_+ + H_-$ with
\begin{eqnarray} \label{decoup}
H_\pm &=& \frac12 \int dx \left \{
(\partial_x \varphi_\pm)^2 + (\partial_x 
\vartheta_\pm)^2 \right\}
\\ \nonumber &&\quad 
\pm \frac{\lambda}{2(\pi a)^2} \cos \left[
\sqrt{8\pi g} \,\vartheta_\pm(0)\right]\;,
\end{eqnarray}
and the boundary conditions (\ref{bc}) 
determining the density $\bar{\rho}_{p,r}$
of $p=\pm$ movers injected into channel $r=\pm$ take the form
\begin{equation}\label{bc2}
\langle \bar{\rho}_{p,r} (x\to - p \infty)\rangle
= \frac{p}{\sqrt{2}} \frac{e (U_1 + r U_2) }{4 \pi g} \;.
\end{equation}
Therefore we are left with two completely decoupled
systems $r=\pm$, each of which is formally identical to the
problem of an elastic potential scatterer embedded
into a spinless LL \cite{kane}. However, this LL now has 
the doubled interaction
strength parameter $\bar{g}=2g$. 
The boundary conditions  (\ref{bc2}) specify  the
effective voltages $\bar{U}_r=
(U_1+r U_2)/\sqrt{2}$ applied to channel $r=\pm$.
In analogy to Eq.~(\ref{boscurr}), currents 
in channel $r=\pm$ are  defined by
$\bar{I}_r=e \sqrt{g/\pi} \, \partial_t
\vartheta_r(0)$, and
from Eq.~(\ref{newbas}), we then  find the currents  
$I_i = (\bar{I}_+ \pm \bar{I}_-)/\sqrt{2}$ flowing
in conductor $i=1,2$.

The Hamiltonian (\ref{decoup}) has been discussed in detail before, see,  
e.g., Refs.\cite{kane,matveev,saleur,guinea,weiss}.
For arbitrary $\bar{g}$, 
the exact solution of the transport problem
has been given in
Ref.\cite{saleur}. This solution exploits 
the integrability of Eq.~(\ref{decoup})
and employs the thermodynamic Bethe ansatz. 
Simpler exact solutions are possible by means of refermionization
techniques
for $\bar{g}=1$ [see Eq.~(\ref{i2})] and $\bar{g}=1/2$.  
The case $g=1/2$ thus corresponds to an uncorrelated  situation 
in the new basis (\ref{newbas}), 
and  $g=1/4$ is the Toulouse point \cite{guinea}.
Progress can also be made by expanding in $|\epsilon|\ll 1$ for 
$\bar{g}=1-\epsilon$ \cite{matveev} or 
$\bar{g}=1/2-\epsilon$ \cite{weiss}.

Employing the exact results of, e.g., Ref.\cite{saleur}, 
at zero temperature we find the asymptotic low-voltage behavior
 \begin{eqnarray}\label{lowvol}
I_i  &=& \frac{e^2}{h} \frac{\lambda_B}{e}\,
 \Bigl \{ \,{\rm sgn}(U_1+U_2) \,
 [e|U_1+U_2|/\lambda_B]^{1/g-1} \\ \nonumber
&& \quad \pm \; {\rm sgn}(U_1-U_2)
\; [e|U_1-U_2|/\lambda_B]^{1/g-1} \Bigr\}\;,
\end{eqnarray}
where the $\pm$ sign corresponds to $i=1,2$, respectively.
The energy scale $\lambda_B$ generated by the 
bare electrostatic coupling $\lambda$  is given by
\begin{equation}\label{scale}
\lambda_B = (c_g/a)\, (\lambda/a)^{1/[1-2g]}\;,
\end{equation}
where $c_g$ is a numerical constant of order unity \cite{saleur}. 
The result (\ref{lowvol})
holds under the condition
\begin{equation}\label{cond}
e|U_1\pm U_2| \ll \lambda_B \;.
\end{equation}
If both voltages approach zero, the linear conductance vanishes
in both 1D conductors.
We thus find a pronounced {\em zero-bias anomaly},
with characteristic interaction-dependent power laws
for small voltages.

Let us now discuss the full current-voltage characteristics.
A particularly simple solution emerges at the Toulouse 
point $g=1/4$ by 
refermionization \cite{guinea} of Eq.~(\ref{decoup}) under the 
boundary conditions (\ref{bc2}).
At zero temperature, the result is
\begin{equation}\label{g14}
\bar{I}_\pm=(e^2/h) \,[U_\pm - V_\pm] \;,
\end{equation}
where $V_\pm$ is the four-terminal voltage \cite{egger96}
subject to the self-consistency equation
\begin{equation}
eV_\pm = 2\lambda_B \tan^{-1}
 \{ [2eU_\pm-(3/2)eV_\pm]/\lambda_B \} \;,
\end{equation}
where $\lambda_B=\lambda^2/4(\pi a)^3$
in accordance with Eq.~(\ref{scale}). 
Under the condition (\ref{cond}), the exact result 
(\ref{g14}) reproduces Eq.~(\ref{lowvol}) again.
In the absence of a coupling, $\lambda_B=0$, one
finds the correct unperturbed currents $I_i=(e^2/h) U_i$.

\begin{figure}
\hfil
\epsfysize=6cm
\epsffile{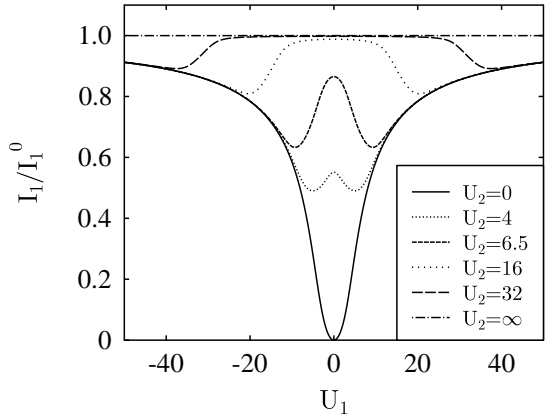}
\hfil
\caption[]{\label{fig2}
Transport current $I_1$ normalized to the unperturbed
value $I_1^0=(e^2/h)\, U_1$
for $g=1/4, \lambda_B/e=1, T=0$, and
several values of the cross voltage $U_2$.}
\end{figure}

The   transport current $I_1$ is plotted 
as a function of $U_1$  in Fig.~\ref{fig2}. 
Contrary to what is found in the uncorrelated 
 system \cite{datta}, the current $I_1$ is extremely sensitive to
the  applied cross voltage $U_2$.  For $U_2=0$, transport
 becomes fully suppressed for $U_1\to 0$, with $g$-dependent
nonlinear low-voltage corrections given by Eq.~(\ref{lowvol}).
Increasing $U_2$ for some fixed $U_1$ then
leads to an increase in the 
current $I_1$. Eventually, the linear conductance behavior 
is restored for very large cross voltage.  
In fact, for $e|U_1\pm U_2|\gg \lambda_B$,
one always recovers the unperturbed currents $I_i=(e^2/h)\,
U_i$.  The generic correlation effects  
 are most important under the conditions (\ref{cond}).

 Remarkably, there is a suppression of the current if $U_1=\pm U_2$,
which is observed as a ``dip''  in the 
normalized current displayed in Fig.~\ref{fig2}.
This effect can be rationalized in terms of 
 a partial dynamical pinning of charge
density waves in conductor 1 due to commensurate
 charge density waves in conductor 2.
As can be checked from Eq.~(\ref{g14}), while the nonlinear
conductance $G_{11}\equiv\partial I_1/\partial U_1$ stays positive,
one can have a negative value for
 $G_{12}\equiv\partial I_1/\partial U_2$.
In fact, the latter (off-diagonal) conductance is within 
 the bounds $-e^2/2h\leq G_{12}\leq e^2/2h$ [we note that
$G_{12}=0$ for $g\geq 1/2$], 
while the diagonal conductance  fulfills
$0\leq G_{11} \leq e^2/h$.
The pronounced and nonlinear  sensitivity of the current $I_1$ 
to the applied cross voltage $U_2$
is a distinct fingerprint for Luttinger liquid behavior.  
Parenthetically, anomalous power laws 
can also be found in the temperature dependence of
the current. 

To conclude, we have examined nonlinear
transport through  two Luttinger liquids
coupled together at one point. 
The only relevant coupling is of electrostatic origin
and leads to distinct correlation effects for strong 
Coulomb interactions, $g<1/2$.
The theoretical findings reported here
could be of use for the experimental identification
of non-Fermi-liquid behavior in carbon nanotubes
and other one-dimensional materials.

We thank A.O. Gogolin, H. Grabert, and C.A. Stafford
for useful discussions. This work was supported by the 
Deutsche Forschungsgemeinschaft (Bonn).


\begin{references}
\bibitem{unp} 
The same effects as discussed here
for spinless electrons  are also found for spin-$\frac12$ electrons
or in carbon nanotubes.
The minor modifications required in these two cases and
the extension to different $g$ parameters for both
conductors will be presented  elsewhere.
\bibitem{fo1}
If the conductors are brought into mechanical
contact, local deformations might cause 
impurity scattering in one or both conductors.
For simplicity, we assume that no such impurity
scattering processes are present.
\bibitem{foot}
For a bulk coupling such as in Refs.\cite{fabrizio,kusmartsev}, a
perturbation is relevant already for $\eta<2$.
\bibitem{fo2} 
If the operators $V_2$ and $V_3$ are taken in a strictly
local sense, the concept of non-zero conformal spin does
not apply and both operators are simply irrelevant. 
In practice, however, they might act over some finite 
(but small) distance around $x=0$. Thereby one generates
the second-order contribution in Eq.~(\ref{vflow}).
\end{references}
\end{document}